\documentclass[aps,prl,twocolumn,superscriptaddress,showpacs,a4paper]{revtex4}
\usepackage{amsfonts}

\usepackage[dvips]{graphicx}
\usepackage[hypertex,breaklinks=true,colorlinks=false]{hyperref}

\usepackage{ifthen}
\usepackage{ulem}

\newcommand{\journal}[4]{\ifthenelse{\equal{#1}{prl}}{\href{http://link.aps.org/abstract/PRL/v#2/e#3}{\prl {\bf #2}, #3 (#4)}}{\ifthenelse{\equal{#1}{prb}}{\href{http://link.aps.org/abstract/PRB/v#2/e#3}{\prb {\bf #2}, #3 (#4)}}{\ifthenelse{\equal{#1}{arxiv}}{preprint \href{http://arxiv.org/abs/#2.#3}{arXiv:#2.#3}}{\ifthenelse{\equal{#1}{rmp}}{\href{http://link.aps.org/abstract/RMP/v#2/e#3}{\rmp {\bf #2}, #3 (#4)}}{\ifthenelse{\equal{#1}{cond-mat}}{preprint \href{http://arxiv.org/abs/cond-mat/#2}{cond-mat/#2}}{\ifthenelse{\equal{#1}{pre}}{\href{http://link.aps.org/abstract/PRE/v#2/e#3}{\pre {\bf #2}, #3 (#4)}}{#1 {\bf #2}, #3 (#4)}}}}}}}

\begin{document}

\title{Correlations and order parameter at a Coulomb-crystal phase transition in a three-dimensional dimer model}

\author{Gr\'egoire Misguich}
\email{gregoire.misguich@cea.fr}
\affiliation{
Institut de Physique Théorique,
CEA, IPhT, CNRS, URA 2306, F-91191 Gif-sur-Yvette, France.}
\author{Vincent Pasquier}
\affiliation{
Institut de Physique Théorique,
CEA, IPhT, CNRS, URA 2306, F-91191 Gif-sur-Yvette, France.}

\author{Fabien Alet}
\affiliation{Laboratoire de Physique Th\'eorique, IRSAMC, Universit\'e Paul Sabatier, CNRS, 31062 Toulouse, France}
 
\begin{abstract}
The three-dimensional classical dimer model with interactions
shows 
an unexpected continuous phase transition between an ordered dimer crystal and
a Coulomb liquid. 
A detailed analysis of the critical dimer and monomer correlation functions
point to a subtle interplay between the fluctuations
of the crystal order parameter and the ``magnetic'' degrees of freedom present in the Coulomb phase.
The distribution probability of the crystal order parameter suggests an
emerging continuous $O(3)$ symmetry at the critical point.
\end{abstract}




\maketitle

{\it Introduction --- } The Berezinskii-Kosterlitz-Thouless transition~\cite{bkt}
is an
example of critical phenomenon
which cannot be understood from the conventional Landau-Ginzburg-Wilson
(LGW) point of view, based on spontaneously broken symmetries
and critical fluctuations of the associated order parameter~\cite{lf80}.
Many quantum systems in one dimension (1D) and classical systems in 2D
exhibit such type of continuous phase transition, which is understood as a proliferation of topological defects.
The recent suggestion \cite{senthil04} that some phase transitions {\it in $D>2$} could also be outside
the LGW framework  triggered an important activity, both on the
field-theoretical~\cite{3dth} and numerical~\cite{mw04,sandvik07,num} sides.

The classical dimer model on the cubic lattice \cite{huse03,alet06} is one
of the simplest 3D models with a transition which does not seem
to fit in the LGW picture \cite{alet06}, although the field theory of the critical point has not yet been elucidated.
Configurations of the model are hard-core dimer coverings of a cubic lattice \cite{huse03}.
Each dimer occupies two sites and each site belongs
exclusively to one dimer. In addition, each configuration has an energy equal to -1 times the number of square plaquettes with two parallel
dimers~\cite{alet06}. At low temperature $T$, this interaction favors the six-fold degenerate
regular (crystalline) arrangements of dimers. At high-$T$, the lattice
symmetries are restored (dimer liquid), but dimer-dimer correlations decay algebraically with distance (with the same $r^{-3}$ dependence as the interaction energy of two magnetic dipoles) \cite{huse03}.
This so-called {\it Coulomb phase} has also been found in other 3D models~\cite{Coulomb-other} and is relevant to some pyrochlore compounds~(see for instance Ref.~\cite{bramwell01-henley05-isakov05}).

The most intriguing properties of this dimer model concern the phase
transition between  the low-$T$ crystal and the Coulomb phase. Previous
simulations~\cite{alet06} indicated that the transition is {\it continuous}
(second order). This is already  in contradiction with the LGW approach, which generically predicts a {\it first order} transition
in the present case. The main question, still open at present,  is to understand what are the fields needed
for a description of the long-distance properties at the critical point, and what is the
associated field theory.
In this Rapid Communication, we report on  numerical investigations of correlation
functions and order parameters in the vicinity of the critical temperature $T_c$.
We show that the dipolar character of the dimer correlations is gradually lost when approaching the transition,
and find an unexpected probability distribution of the order parameter, with a possible
emerging {\it continuous} symmetry.
Related to rotations of a 3-component order parameter, this $O(3)$ symmetry is not present at the microscopic level. Terms which make the symmetry group discrete at the lattice scale
(a dimer can only take six orientations) may be irrelevant in the
renormalization group (RG) sense at the critical point,
but clearly relevant in the ordered phase. This finding
is a step toward a description of the transition in the continuum limit, and establishes 
a tight connection with previous models (quantum \cite{sandvik07} and classical \cite{mw04})
discussed in the context of non-LGW transitions.

\begin{figure}
\hspace*{-0.7cm}
\includegraphics[width=4.7cm]{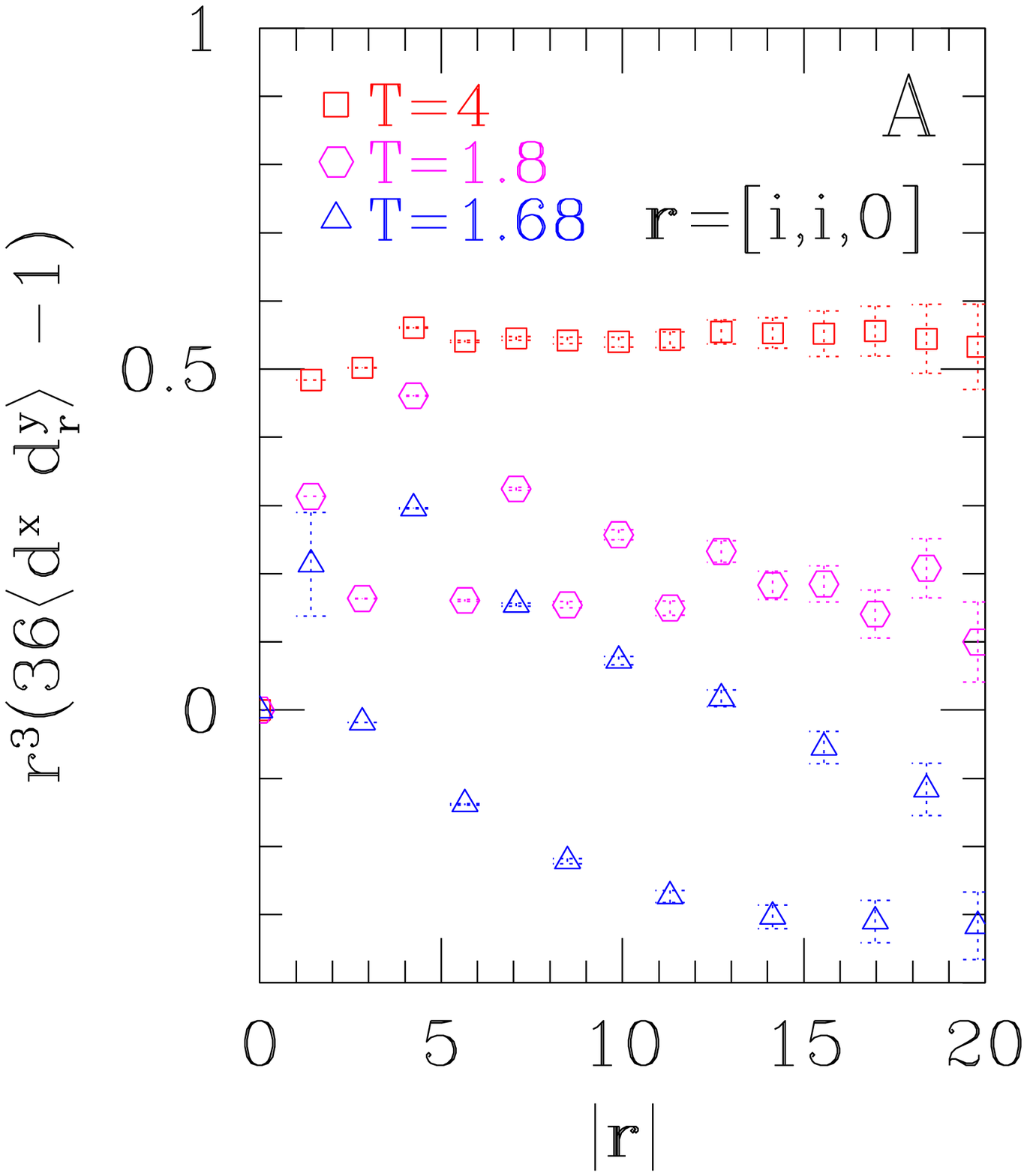}\hspace*{-0.6cm}
\includegraphics[width=4.7cm]{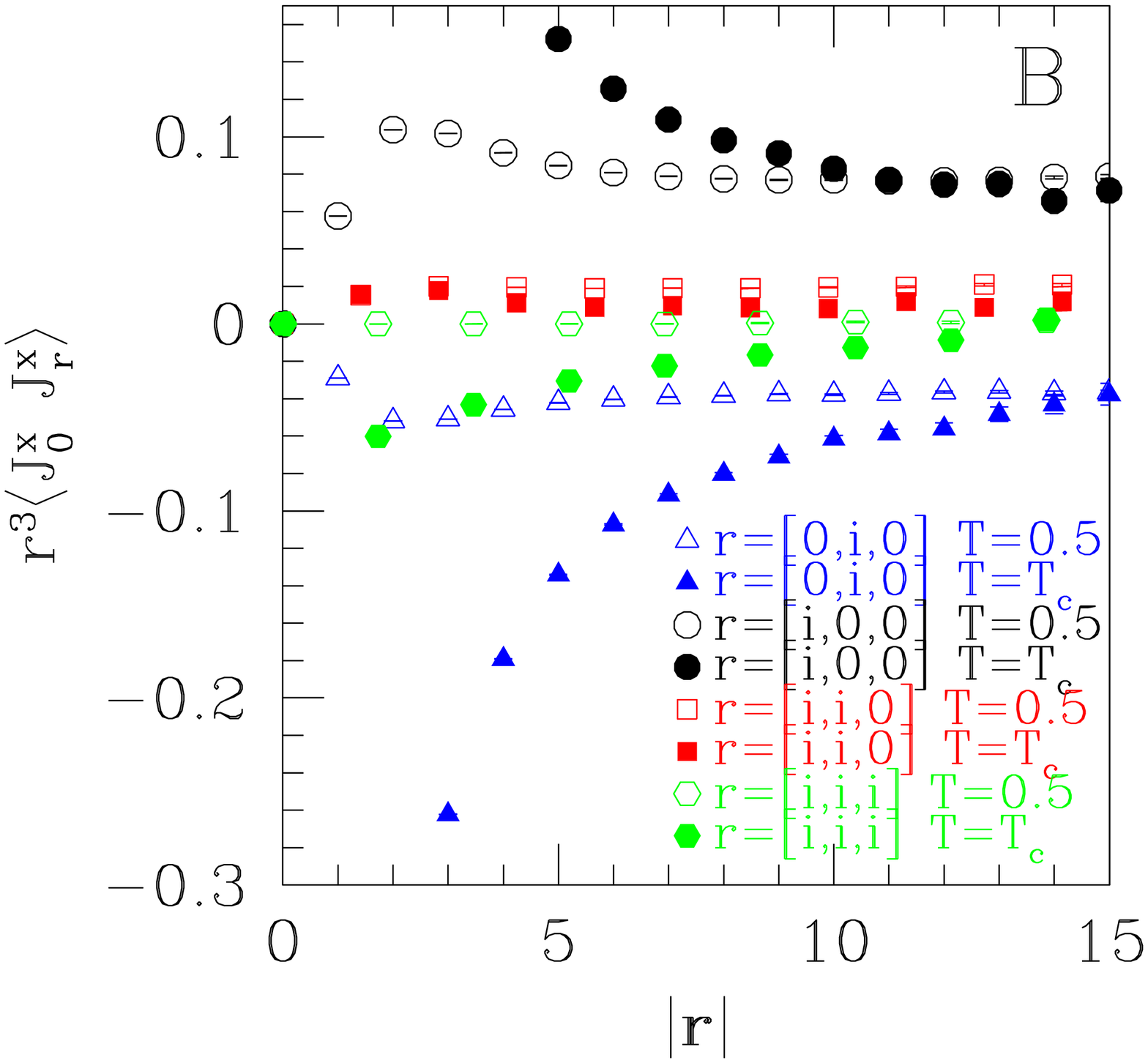}\vspace*{-0.5cm}
\caption{(color online) 
A: Dimer correlations  ${\langle}d^x_{\bf 0} d^y_{\bf r}{\rangle}$
as a function of $|{\bf r}|$ and multiplied by $|{\bf  r}|^3$ 
for a few selected $T$ above $T_c$ (system size  $L=64$).
For this orientation, the {\it spin} contribution vanishes.
B: Current correlations  ${\langle}J^x_{\bf 0} J^x_{\bf r}{\rangle}$ of the
ICLM as a function of $|{\bf r}|$ and multiplied by $|{\bf  r}|^3$, 
at $T=0.5$ (in the Coulomb phase) and $T_c=0.33305$ (system size  $L=64$). Correlators at $T_c$ have been multiplied by 50.
Simulations of the dimer model (ICLM) were performed with the algorithm of Ref.~\onlinecite{sandvik06} (Ref. ~\onlinecite{as03}).
}
\label{fig:dip-jj}
\end{figure}

\begin{figure}
\includegraphics[width=8cm]{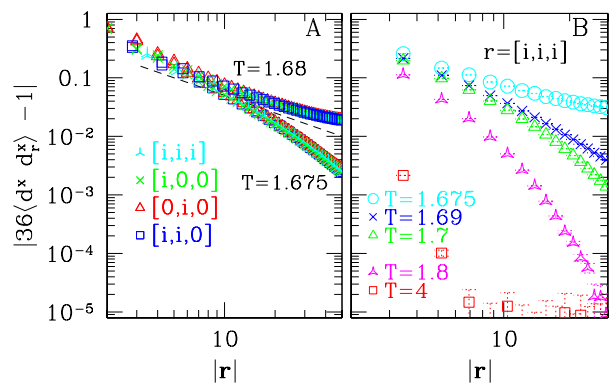}\vspace*{-3.2cm}
\caption{(color online) 
Dimer  correlations  ${\langle}d^x_{\bf 0} d^x_{\bf r}{\rangle}^c$.
A: $T\simeq T_c$ and just above, for different orientations of $\bf r$ ($L=128$).
B: A few selected $T$ above $T_c$ ($L=64$).
The lattice direction ${\bf r}=[i,i,i]$
is such that the {\it dipolar} contribution vanishes.
 }
\label{fig:spin}
\label{fig:iso}
\end{figure}

{\it Dimer correlations: from dipolar to spin-spin form --- } 
In Coulomb phases, it is useful to introduce a  ``magnetic field'' \cite{huse03}
defined on each bond 
by $B_{ij}=-B_{ji}=1$
if a dimer sits between sites $i\in A$ and $j\in B$ ($A$ and $B$ are the two sublattices),
and $B_{ij}=0$ otherwise.
When considered as a vector field ${\bf B}$, its lattice divergence satisfies ${\bf \nabla} \cdot {\bf B} =  (-)^{\bf r} =  \pm 1$, as each
site is occupied by exactly one dimer.
Coarsegraining over a large number of sites, the contributions of both sublattices
cancel out to give ${\bf \nabla} \cdot {\bf B} =0$.
In the continuum limit, ${\bf B}$ is thus written using a vector potential:
${\bf B}={\bf \nabla}\times{\bf A}$.
At high-$T$ and long distances, this zero-divergence constraint
captures the essential aspects of the dimer hardcoreness~\cite{huse03}.
The Coulomb phase is then described by the simplest action compatible with the gauge invariance
${\bf A}\to  {\bf A}+{\bf \nabla} \Lambda$ ($\Lambda$ is an arbitrary function), that is
$S=\frac{K}{2}\int d^2{\bf r}  ({\bf \nabla}\times{\bf A})^2({\bf r})$. Here the rigidity $K^{-1}$ depends on $T$.
This effective action
leads to  dipolar long-distance correlations~\cite{huse03}:
\begin{equation}
 \langle
 	B^\alpha(0) B^\beta({\bf r})
 \rangle^c
 \sim \frac{1}{K(T)}
 	\frac{3 r^\alpha r^\beta-\delta^{\alpha\beta}{\bf r}^2}{{\bf r}^5}.
\label{eq:dipolar}
\end{equation}
$K^{-1}$ can be measured numerically through $L K^{-1}=\langle (\phi^z)^2 \rangle$, the fluctuations of the total flux $\phi^z=\int dxdy B^z({\bf r}=[x,y,z_0])$ going through the sample
 (independent of $z_0$).
Ref.~\onlinecite{alet06} showed that $K^{-1}$ 
{\it vanishes continuously when approaching the transition} at $T_c\simeq 1.675$, as 
magnetic field fluctuations are costly when approaching the crystal.
Although the dipolar form dominates 
at high-$T$, we find that this contribution to the dimer-dimer correlations (Eq.~\ref{eq:dipolar})
progressively disappears when approaching $T_c$. 
In Fig.~\ref{fig:dip-jj}A, the correlations $\langle d^x_0 d^y_{\bf r}\rangle^c$
for ${\bf r}=[i,i,0]$ 
 are multiplied by $|{\bf r}|^3$ to highlight the dipolar contribution,
 and its strong reduction when going from $T=4$ to $T=1.68$ (just above $T_c$).
This is expected since $K^{-1}$ vanishes when approaching $T_c$. 
However, the ``magnetic'' action
does not account for an important qualitative feature.
Setting $\alpha=\beta=x$ in Eq.~\ref{eq:dipolar}, the prefactor $a$ of the dipolar contribution $a/r^3$,
depends on the lattice direction (for instance $a=0$ for ${\bf r}=[i,i,i]$
 and $a=1/2$ for ${\bf r}=[i,i,0]$), whereas the actual
 correlations close to $T_c$ appear to be almost {\it isotropic} (see Fig.~\ref{fig:iso}A).

Eq.~\ref{eq:dipolar} misses an important physical ingredient: the 
crystal-like dimer correlations associated to the order parameter of the symmetry breaking phase. 
At each site, we define a three-component vector $\vec n$
by $n^\alpha({\bf r})=\pm (-1)^{r_\alpha}$ if the dimer
sitting in ${\bf r}$ points in the direction $\pm \alpha$, and 0 otherwise.
This insures that $\vec n$ is ``ferromagnetically'' correlated in the crystal phase 
($\langle \vec n\rangle=\pm \vec{e_x},\pm \vec{e_x},\pm \vec{e_y}$ in the six ground-states).
Making the assumption that long-distance fluctuations of  $\vec n$
are described by a three-component ``spin'' model ({\it  i.e.} absence
of coupling between real space $r^\alpha$ and internal spin $n^i$ indices, as in
$O(n)$ models),
the angular dependence of the correlations become
$
\langle n^\alpha(0) n^\beta({\bf r}) \rangle^c\sim \delta^{\alpha\beta}
\label{eq:O3correl}
$.
It turns out that in a wide temperature range around $T_c$,
the connected dimer-dimer correlations $\langle d^\alpha(0) d^\beta({\bf r})) \rangle^c$ are well approximated
by a dipolar contribution, plus a ``spin-like'' contribution:
\begin{equation}
  		\frac{(-1)^{\bf r}}{K}
 		\frac{3 r^\alpha r^\beta-\delta^{\alpha\beta}{\bf r}^2}{{\bf r}^5}
 		+(-1)^{r_\alpha}g(|{\bf r}|)\delta^{\alpha\beta}.
\end{equation}

Taking  $\alpha=x,\beta=y$ and ${\bf r}=[i,i,0]$ (Fig.~\ref{fig:dip-jj}A), the ``spin'' term vanishes
and only the dipolar part remains, whereas for
$\alpha=\beta=x$ and ${\bf r}=[i,i,i]$  the
dipolar part vanishes and one is left with $g(r)$.
The latter is displayed in Fig.~\ref{fig:spin}B, where strong and slowly decaying correlations are found
close to $T_c$.
At $T_c$, $g(r)$ is very likely algebraic
and compatible with $\sim 1/r$, although it is difficult to give an
error bar on the exponent.

\begin{figure}
\includegraphics[width=7.5cm]{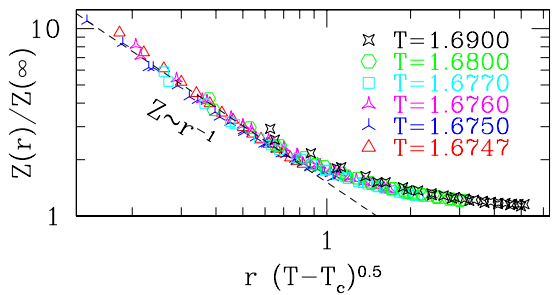}\vspace*{-3.3cm}
\caption{(color online) 
Monomer correlation
as a function of $r/\xi$ in log-log  scale
for a few $T$ just above $T_c$, assuming
$\xi=(T-T_c)^{-\frac{1}{2}}$ and  $T_c\simeq 1.6743$ ($L=128$).
 }
\label{fig:mono}
\end{figure}

A more quantitative analysis can be done on the monomer correlator $Z(r,T)$,
defined as the ratio of the partition function with two fixed test monomers at distance $r$, by the
bare partition function.
In the Coulomb phase, $Z(r,T)$ converges to a finite value when $r\to\infty$ \cite{alet06}.
As shown in Fig.~\ref{fig:mono}, $Z(r,T)/Z(\infty,T)$ is well fitted
by a function $F(x)$ of the single parameter $x=r\sqrt{T-T_c}$, suggesting
the existence of a length $\xi(T)\simeq (T-T_c)^{-\frac{1}{2}}$, diverging  at $T_c$
with an exponent $\nu$ compatible with $0.5$. 
By construction $F(x\to\infty)=1$, and we observe  $F(x)\sim 1/x$ for $x\lesssim 1$, indicating
that for distances $r \lesssim \xi(T)$, $Z(r,T)$ decays in a way consistent with  $\sim 1/r$.
At distances large compared to the lattice spacing, but small compared to some
(diverging) correlation length $\xi(T)$, the correlations
 are thus qualitatively {\it not} of Coulomb type, although $T>T_c$.
Below  $\xi$, the correlations (of both dimers and monomers) are compatible
with an exponent $\eta$ close to zero. Noteworthy, these values are those
obtained from independent thermodynamic measurements \cite{alet06}.

It is instructive to compare these results with the correlations  of a simpler model:
the ``integer current loop model'' (ICLM)~\cite{wsgy94,as03}. There,  the allowed configurations are
integer-valued currents $J_{ij}=-J_{ji}\in\mathbb{Z}$ defined on each bond $(ij)$ of the cubic lattice
(analogous to the dimer magnetic field ${\bf B}$), with the constraint
that the net current arriving at each site is zero: ${\rm div} \vec J=0$.
The energy of each configuration is $E=\frac{1}{2}\sum_{\langle i,j\rangle}{(J_{ij})}^2$.
This model also has a Coulomb phase at high-$T$, with large current fluctuations and dipolar
$\langle J^\alpha J^\beta \rangle$
correlations.
At low-$T$, small current loops dominate and
eventually $J_{ij}=0$ everywhere at $T=0$ (a state which does
{\it not} have a dimer analog).
Through a standard duality transformation~\cite{jkkn77}, the ICLM maps exactly onto an
$O(2)$ spin model with Villain interaction~\cite{villain75}. In contrast to the dimer model, the phase transition of the ICLM ($T_c^{\rm ICLM}\simeq0.33305$~\cite{as03}) is known to belong to the 3D-XY universality class.
The Coulomb phase of the ICLM corresponds to the ordered phase in the
angular variables $\theta_{\bf r}$, and the dipolar correlations are a
direct consequence of the Goldstone mode for these dual spins. At high-$T$, this dictionary works also well for the dimer model:  $K^{-1}$ is the {\it stiffness} of a dual $O(2)$ ferromagnet. The monomer correlation $Z(r)$ corresponds to the spin-spin correlation $\langle \exp{(i\theta_0-i\theta_r)}\rangle$,
and $Z(\infty)$ (finite in the Coulomb phase and vanishing at $T_c$)
to the square of the order parameter (magnetization) $|\langle e^{i\theta_0} \rangle|^2$.
The crucial 
difference with  the dimer model is the absence of spontaneously broken
symmetry in the low-$T$ phase of the ICLM,
and thus
the absence of an order parameter similar to $\vec n$. This does not only make the critical exponents of the dimer
model different from that of 3D XY~\cite{alet06}, but marks a {\it qualitative} difference concerning the correlation in the Coulomb phase close to $T_c$ (for $r\lesssim \xi(T)$).
The current correlations of the ICLM {\it remain dipolar down to $T_c$}.
Both in the Coulomb phase and at $T_c$, they decay as $\sim 1/r^3$ with a geometric prefactor which does depend on the direction of $\bf r$
(see the large-$|\bf r|$ behavior in Fig.~\ref{fig:dip-jj}B).
The ICLM obeys  Eq.~\ref{eq:dipolar} without the need for 
an additional ``spin like'' contribution.
Only the overall amplitude is reduced close to $T_c$, with $1/K(T_c)\sim 1/L$
(see renormalization factor in Fig.~\ref{fig:dip-jj}A).
The fact that dimer
correlations are dominated by non-dipolar contributions close to $T_c$
gives additional support for this transition not being of 3D XY nature, as
naively expected from the high-$T$ side.

\begin{figure}
\hspace*{-0.3cm}
\includegraphics[width=6cm,angle=-90]{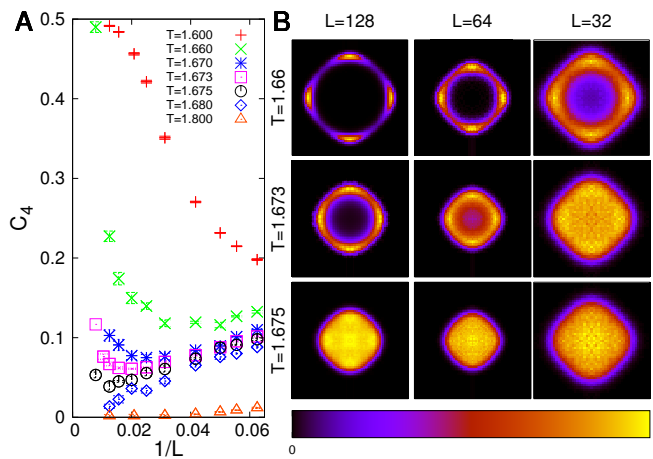}
\vspace*{0.2cm}
\caption{(color online) 
A: 
Spherical  character  of   the
order parameter  $\vec     N$ distribution ($C_4$) as a function of $1/L$ for different $T$.
The data for $T=1.6$  have been divided by 2.
B: Probability distribution $P(N^x,N^y,|N^z|\leq 0.03)$ as a function of
$N^x$ and $N^y$ for several temperature and system sizes.
The range is $N^{x,y}\in[-0.3,0.3]$ for $L<128$ and $[-0.2,0.2]$
for $L=128$. 
}
\label{fig:cos4}
\label{fig:distribN}
\end{figure}

{\it Symmetry in the crystal order parameter at $T_c$ --- }
The results above point to the importance of fluctuations of the crystal order parameter $\vec n$ at the transition. We therefore looked at the distribution probability
of $\vec N=L^{-3}\sum_{\bf r} \vec n({\bf r})$ close to $T_c$.
For large systems,
 the probability distribution of $(N^x,N^y,N^z~\simeq 0)$ is almost circular (see
 the distribution for $L=128$ and $T=1.673$ in
 Fig.~\ref{fig:distribN}B). Some small square-like anisotropy is still
 detectable but we argue  below that it could be a finite size effect and the asymptotic
 form of the distribution spherically symmetric in
   the thermodynamic limit. To quantify this, we define
\begin{eqnarray}
	C_{4}&=&\frac{1}{2}
		\langle
			\cos(4\arg[N^x+iN^y])
			+\cos(4\arg[N^y+iN^z])\nonumber\\
			&&+\cos(4\arg[N^z+iN^x])
		\rangle,
\label{eq:cos4}
\end{eqnarray}
which measures the $O(3)$ character of the angular
distribution of $\vec N$.  If  $\vec N$ is distributed  isotropically,
$C_{4}$  vanishes in the thermodynamic limit. If $\vec N$ preferentially points
along  the  axis 
directions, one expects $C_{4}>0$.
In the crystal phase $C_4\to 1$ when $L\to\infty$, signaling that the direction of the order
parameter $\vec N$ is locked along one crystal axis
(see
 the $C_4(T=1.66)$ and the associated distributions in
 Fig.~\ref{fig:distribN}A and B). In the Coulomb phase,
$\vec N$ is typically very small ($\langle \vec N^2\rangle\sim L^{-3}$)
and distributed in a Gaussian way, thus with spherical symmetry.
The non-trivial result is the behavior of $C_4$ {\it at} $T_c$.
Fig.~\ref{fig:cos4}A suggests that $C_4(T=T_c,L)\to 0$ when $L\to\infty$,
and thus indicate an $O(3)$ symmetric distribution of the order parameter~\cite{foot}.

For $T$ slightly below $T_c$, Fig.~\ref{fig:cos4}A reveals that $C_4(T,L)$ has a minimum for a system size $L_0(T)$.
This minimum shifts to larger size when approaching $T_c$ from below.
$L_0(T<T_c)$ thus defines a lengthscale, diverging at $T_c$, up to which the discrete symmetry breaking
becomes weaker at long distances.

This is similar to the 3D XY model with $Z_{q\geq4}$ anisotropy~\cite{jkkn77,lsb07},
where the $q-$fold anisotropy is dangerously irrelevant and makes the (2-component) order parameter apparently $U(1)$ symmetric in the ordered phase, up to some lengthscale diverging at $T_c$.
A crucial difference here is that the order parameter $\vec n$ has 3 components, and a RG analysis
of the $O(3)$ model predicts that cubic anisotropies, such as
$(n^x)^4+(n^y)^4+(n^z)^4$, should be {\it relevant}~\cite{carmona00}.
The latter anisotropic $O(3)$ model is however qualitatively  different
from the model studied here, as its high-$T$ phase is a featureless paramagnet.
From this point of view, the classical spin model of Ref.~\onlinecite{mw04} (where a 
constraint equivalent to ${\rm div} {\bf B}=0$ is enforced on spin configurations) may be 
closer to our study since it also possesses a high-$T$ Coulomb phase. Discrete anisotropies 
seeming irrelevant for the dimer critical point, both models would be predicted from the 
symmetry point of view to share the same universality class. However, based on the values of 
the critical exponents, this is not the case.

{\it Discussion --- } The numerical observations of this Rapid Communication emphasize
the unconventional nature of the Coulomb-crystal phase transition. 
On one side, the presence of a diverging lengthscale below which correlations
are not Coulomb-like cannot be interpreted by the
sole high-$T$ ``magnetic'' point of view. From the ordered side, we exhibited a lengthscale below which the crystal order
parameter  seems to fluctuates in an unexpected $O(3)$ symmetric way. Besides the difficulty to build
a LGW theory from the corresponding order parameters~\cite{alet06}, those are
clear qualitative indications that individual (magnetic or crystalline)
descriptions cannot capture the physics involved at the transition.
This points to the existence of a critical regime where the interaction
between the ``magnetic'' degrees of freedom at play in the Coulomb phase and
the ``spin'' degrees of freedom which order in the crystal dominates the long distance physics. 
This phenomenology is similar to that of a non-compact CP$^1$ (NCCP1) model \cite{mw04,senthil04}, which precisely
couples spin and gauge degrees of freedom. The NCCP1 theory is characterized
by two continuous symmetries: $SU(2)$ for spin rotations, and $U(1)$
associated to the magnetic flux conservation. In our case, the flux
conservation is implemented at the lattice level, and we argued that
continuous ``spin'' rotations emerge at the critical point. This is similar
to the 2D quantum spin system studied in Ref.~\onlinecite{sandvik07}, but there the spin rotation symmetry is
explicit and a gauge flux conservation is argued to emerge dynamically at the critical point.
Despite all these common ingredients between NCCP1 theories and our dimer model, 
there is a clear mismatch between our
critical exponents and those attributed so far to NCCP1 critical
points~\cite{mw04,sandvik07}.
So, the actual field theory describing the dimer model transition remains
to be uncovered.
Besides its
theoretical importance, the issue of the universality class of this
transition will also become experimentally relevant if a Coulomb-crystal phase transitions was observed in nature,
as in a frustrated (ice-like) magnet for instance.

{\it Acknowledgments --- } G.M. acknowledges K.~S. Kim for useful discusions.
Calculations were performed at CCRT/CEA (project 575), using the ALPS
libraries~\cite{ALPS}.


\begin{thebibliography}{99}

\bibitem{bkt} V. L. Berezinskii, Sov. Phys. JETP  {\bf 32}, 493 (1971);
J.M. Kosterlitz and D.J. Thouless, J. Phys. C {\bf 6},  1181 (1973);
J. M. Kosterlitz, J. Phys. C. {\bf 7}, 1046 (1974) 

\bibitem{lf80} See {\it e.g.} L.D. Landau and E.M. Lifshitz, {\it
  Statistical Physics}  3rd ed.,  (Pergamon, New York, 1980).

\bibitem{senthil04} T. Senthil {\it et al.}, Science {\bf 303}, 1490 (2004);
  \journal{prb}{70}{144407}{2004}.

\bibitem{3dth} D.L. Bergman, G.A. Fiete and L. Balents,
\journal{prb}{73}{134402}{2006};  O.I. Motrunich and T. Senthil,
{\it ibid} \href{http://link.aps.org/abstract/PRB/v71/e125102}{{\bf 71}, 125102 (2005)};
P. Ghaemi and T. Senthil, {\it ibid} \href{http://link.aps.org/abstract/PRB/v73/e054415}{{\bf 73}, 054415 (2006)};
K. Gregor and O.I. Motrunich, {\it ibid} \href{http://link.aps.org/abstract/PRB/v76/e174404}{{\bf 76}, 174404 (2007)};
T. Grover and T. Senthil, \journal{prl}{98}{247202}{2007};
O.~I.~Motrunich and  A.~Vishwanath, \journal{arxiv}{0805}{1494}{}.
\bibitem{mw04} O.I. Motrunich and A. Vishwanath,
\journal{prb}{70}{075104}{2004}.

\bibitem{sandvik07}
A.W. Sandvik, \journal{prl}{98}{227202}{2007};
R.G. Melko and R.K. Kaul, {\it ibid}
\href{http://link.aps.org/abstract/PRL/v100/e017203}{{\bf 100}, 017203 (2008)}.

\bibitem{num}
A. Kuklov, N. Prokof'ev and B. Svistunov, 
\journal{prl}{93}{230402}{2004}.

\bibitem{huse03} D. Huse {\it et al.},
\journal{prl}{91}{167004}{2003}.

\bibitem{alet06} F. Alet {\it et al.},
\journal{prl}{97}{030403}{2006}.

\bibitem{Coulomb-other} J. Villain, Solid State Commun. {\bf 10}, 967
  (1972); S.V. Isakov {\it et al.}, \journal{prl}{93}{167204}{2004};
  A. Banerjee {\it et al.}, \journal{prl}{100}{047208}{2008};
  T.S. Pickles, T.E. Saunders and J.T. Chalker,   \journal{arxiv}{0708}{3791}{×}.
  
  
\bibitem{bramwell01-henley05-isakov05}
S. Bramwell {\it et al.},
\journal{prl}{87}{047205}{2001};
C.L. Henley, \journal{prb}{71}{014424}{2005};
S. V. Isakov {\it et al.}, \journal{prl}{95}{217201}{2005}.

\bibitem{sandvik06}
A.W. Sandvik and R. Moessner, \journal{prb}{73}{144504}{2006}.

\bibitem{wsgy94}
M. Wallin  {\it et al.}, \journal{prb}{49}{12115}{1994}.

\bibitem{as03}
F. Alet and E.S. S\o rensen, \journal{pre}{67}{015701}{2003}.

\bibitem{jkkn77}
J. V. Jos\'e {\it et al.}, \journal{prb}{16}{1217}{1977}.

\bibitem{villain75}
J. Villain, J. Phys. (Paris) {\bf 36}, 581 (1975).

\bibitem{foot}Since $T=1.675$ may be slightly {\it below} the actual $T_c$,
$C_{4}(T=1.675)$ may start to increase for $L\geq 128$.

\bibitem{lsb07}
J. Lou, A. W. Sandvik and  L. Balents, \journal{prl}{99}{207203}{2007}.

\bibitem{carmona00}
J. M. Carmona  A. Pelissetto, and E. Vicari, \journal{prb}{61}{15136}{2000}.

\bibitem{ALPS} F. Albuquerque {\it et al.}, J. Magn. Magn. Mater. {\bf 310},
  1187 (2007); M. Troyer, B. Ammon and E. Heeb, Lect. Notes Comput. Sci., {\bf
  1505}, 191 (1998). See {\tt \href{http://alps.comp-phys.org}{http://alps.comp-phys.org}}.

\end{thebibliography}
\end{document}